# Angular dependence of direct rf power absorption studies in Bi$_2$Sr$_2$CaCu$_2$O$_8$ single crystals

S. P. Chockalingam, S. Sarangi and S. V. Bhat [1]

Department of Physics, Indian Institute of Science, Bangalore-560012, India

K. Oka * and Y. Nishihara **

* National Institute of Advanced Industrial Science and Technology, Tsukuba, 3108152, Japan

** Faculty of Science, Ibraki University, Mito 3108152, Japan

[1]Corresponding author:

Prof. S.V.Bhat

Department of Physics

Indian Institute of Science

Bangalore –560012, India

e-mail: svbhat@physics.iisc.ernet.in

Tel.: +91-80-22932727. Fax: +91-80-23602602

**Abstract:**

We have studied rf power dissipation in $Bi_2Sr_2CaCu_2O_8$ single crystals as a function of angle θ between the applied field and c-axis of the crystal by using a novel technique of measuring direct rf power absorption. We observed two peaks in power absorption as a function of temperature. The amplitude and position of the first peak (peak A) change with orientation and attributed to Josephson junctions decoupling. The second peak (peak B) appears just below $T_c$ has been attributed to the vortex motion and its amplitude depends on the angle of orientation.

# Introduction:

Energy dissipation in superconductors is the subject of interest for long time [1]. Angle dependent studies provide the better understanding on the effects of anisotropy behavior of the sample in the power dissipation. The main mechanisms for the ac power dissipation in high $T_c$ superconductors are Josephson junctions (JJ) decoupling (which are invariably present in the high $T_c$ materials, either as extrinsic junctions due to intergranular contacts or as intrinsic junctions in strongly layered superconductors [2] like $Bi_2Sr_2CaCu_2O_8$) and due to Lorentz force driven motion of the vortex lines (Josephson and pancake) in these strongly type II superconductors [3]. The dissipation studies are carried out through the resistivity measurements [4], I-V characteristics [5] power absorption [6] and others.

Many reports are available about power dissipation at microwave frequencies [6] but at radio frequencies only few reports are available [6], so understanding the behavior of superconductors at rf becomes important. In the power absorption studies the amount of power absorbed by the sample is equal to the amount of dissipation and most of the earlier studies have used the technique of non-resonant microwave absorption [NRMA] or non-resonant rf absorption [NRRA] where only the derivative of the power absorbed dP/dH is measured. For the better understanding of dissipation due to various mechanisms and to get clear information on the material characteristics, which is crucial for applications, we carried out detailed studies of direct power absorption using a novel technique [7] in the single crystals of $YBa_2Cu_3O_{7-\delta}$ [8], $La_{2-x}Sr_xCuO_4$ [9] and $Bi_2Sr_2CaCu_2O_8$ [10]. In the present work we report the results of direct rf power absorption from the well characterized single crystals of $Bi_2Sr_2CaCu_2O_8$ as a function of temperature and the angle θ between the applied field and c-axis and explained it in terms of JJ decoupling model [10,11] and in terms of vortex motion.

**Experimental**:

Air annealed single crystals of $Bi_2Sr_2CaCu_2O_8$ grown by traveling zone flux method were used for our studies. The crystals had a nominal $T_c$ of 85K with the transition width t=1K as determined by ac susceptibility measurements. A customized low temperature insert is used to integrate the experiment with a commercial oxford instruments cryostat and temperature controller (4.2K<T<300K) along with a Bruker electromagnet (0<H<1.4T). The absolute power absorption studies are carried out using an rf oscillator designed and fabricated in our laboratory [7]. The system consists of a self-resonant LC tank circuit driven by a NOT gate. The samples under investigation are placed in the core of an inductive coil forming the LC circuit and the power absorption is determined from the measured change in total current supplied to the oscillator. The sample assembly could be rotated around vertical axis so as to vary the angle θ between the magnetic field and c-axis of the $Bi_2Sr_2CaCu_2O_8$ single crystal. Power absorption was measured as a function of temperature and magnetic field applied at different angles.

**Results and Discussions:**

The absolute rf power absorbed by the sample as a function of temperature in the presence of magnetic field of 6000 gauss applied at different angles θ between the field and c-axis of the crystal is shown in the fig-1. As expected in the normal state the power absorption decreases linearly with temperature like the resitivity behavior of metals. The sample shows the transition to superconducting state at $T_c$=85K, in the superconducting state we observed two peaks (peak A and peak B) in power absorption as a function of temperature. Peak A and B overlapped at the point P as in fig.1 which is consider as the minimum of both peaks and the peaks amplitude is calculated from that point P. In the superconducting state the power dissipation is more than in the normal state, which is due to the decoupling of intrinsic josephson junctions. When the induced rf current exceeds the critical current corresponds to the coupling energy of $E_j$ (=$\Phi I_c / 2\pi$) of the josephson junctions then they absorb power and decoupled to normal state where the normal state loss also leads to dissipation. In the sample there are many josephson junctions with different coupling energy, so throughout the temperature some junctions decoupled and dissipates power, for details refer our earlier reports [10,11].

Peak A arises because of decoupling of josephson junctions as explained above. Results of fig-1 shows that with decreasing temperature the number of JJ increases and its coupling energy also goes on increasing, so absorbed power also increases and at certain temperature when critical current goes more than the applied rf current ($I_c > I_{rf}$) then $I_{rf}$ is not able to decouple the junctions and therefore a decrease in the absorbed power is observed. Amplitude of the peak A increases when the angle between the applied field and c-axis of the sample changing from 0° to 90° and there was an abrupt increase in power absorption at 90° as shown in the fig-1 and 2. This behavior is explained as follows the normal state resistance of the $Bi_2Sr_2CaCu_2O_8$ single crystal is large along the c-axis and small across a-b plane. According to Ambegaokar-Baratoff relation the critical current density of josephson junction is inversely proportional to normal state resistance [12] so when the field is perpendicular to c-axis $I_c$ is large and when the field is parallel to c-axis $I_c$ is less. Also as discussed in earlier reports [13,14,15] single crystals are made up of stacking of strong and weak superconducting layers. When the field is applied perpendicular to c-axis then the flux lines are injected parallel to the CuO layers and strongly pinned in the weak superconducting layers, so the critical current density is high at this orientation and decreases at certain fashion in other orientations as in the above reports. If the critical current density is high than the JJ absorb more power to decouple, therefore the power absorption is more. When the direction of the magnetic field is slightly tilted from the layers, then the critical current density drastically decreases [16], so that we observed a abrupt decrease in power absorption when the orientation changes from 90° as shown in the fig.2. Thereafter when the orientation changes towards 0° the pinning in the weak superconduting layers slowly decreases and thus the critical current and the absorbed power when the field is parallel to c-axis the flux pinning is less and JJ needs less power to decouple. We observe a shift in peak A position towards high temperature as shown in the fig1 and fig.2.This shift indicates that critical current $I_c$ increases when the orientation changes from 0° to 90°, so that $I_{rf}$ is not able to decouple JJ at higher temperature and the power absorption starts decreasing as discussed earlier.

Another peak in absorption peak B appears just below $T_c$ and this arises duo the motion of vortices [2]. At higher temperature vortices are not strongly pinned and they leads to dissipation due to lorentz force driven motion, with decreasing temperature the vortices slowly get pinned and dissipation due to its motion starts decreasing .At certain temperature all the vortices get pinned and there is no dissipation, this behavior of vortices leads to a peak in absorption as a function of temperature as shown in fig.1.

We observed a maximum amplitude of peak when the field is parallel to c-axis and when the field is perpendicular the peak vanished as shown in the fig.1 this indicates that in this studies we observed the dissipation which is due to pancake vortices, which can excite when the field is parallel to c-axis and the number of pancake vortices varies as a cosine function of the applied magnetic field as discussed in many reports earlier [2]. The peak position is independent of orientation as in fig.1 indicates that only at higher temperature the vortices get depinned. When the orientation changes from 0˚ to 90˚ the amplitude of the peak decreasing as shown in the fig.1 and fig.4. The amplitude of peak B is plotted in the fig.4 along with the calculated value of dissipation due to the pancake vortices whose number varies as a cosine function of the applied field, the experimental and calculated values are coincided which confirms that dissipation is due to pancake vortices only and the consideration that point P in fig.1 is the minimum of both peaks from where peaks amplitude are calculated is justified. The peak position is independent of orientation indicates that only at higher temperature the vortices get depinned.

## Summary:

The direct rf power absorption studies reveals that dissipation arises due to JJ decoupling and Lortenz force driven motion of pancake vortices. With the change in orientation the critical current density corresponds to decoupling energy $E_j$ of the JJ varies and thus the power dissipation. The vortices contribution in dissipation decreases when the angle changes from 0˚ to 90˚ and vanished at 90˚. We provide a qualitatively satisfactory explanation for this behavior of power dissipation in $Bi_2Sr_2CaCu_2O_8$ single crystals as function of orientation of the applied field.

**Acknowledgements**, SVB thanks the University Grants Commission, India for funding this work.

**Figure Captions:**

1. The absolute rf power dissipated by the sample as a function of temperature in the presence of magnetic field of 6000 gauss applied at different orientations θ between the field and c-axis of the crystal. To be noted are the increase in the amplitude and shift in position of the peak A and the decrease in the amplitude of the peak B.

2. Peak A amplitude as a function of the angle θ between the field and c-axis. The amplitude increases when θ changes from 0° to 90°degrees and there is abrupt increase in amplitude for 90° orientation.

3. Angle dependence of peak A position in temperature scale, when θ changes from 0° to 90° degree the peak is shifted towards high temperature and the shift is more for 90° degree orientation.

4. Peak B amplitude as a function of the angle θ between the field and c-axis along with the calculated value of dissipation due to the pancake vortices whose number varies as a cosine function of the applied field, the experimental and calculated values are coincided. The amplitude decreases when θ changes from 0° to 90° degrees and the peak vanished for 90° degree orientation.

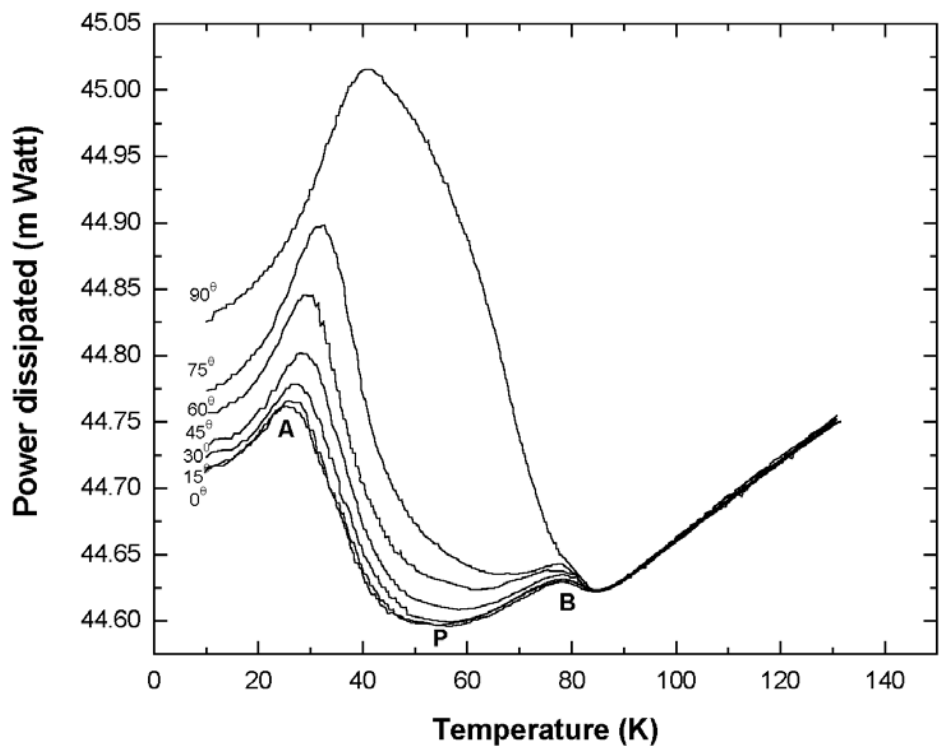

**Fig. 1**

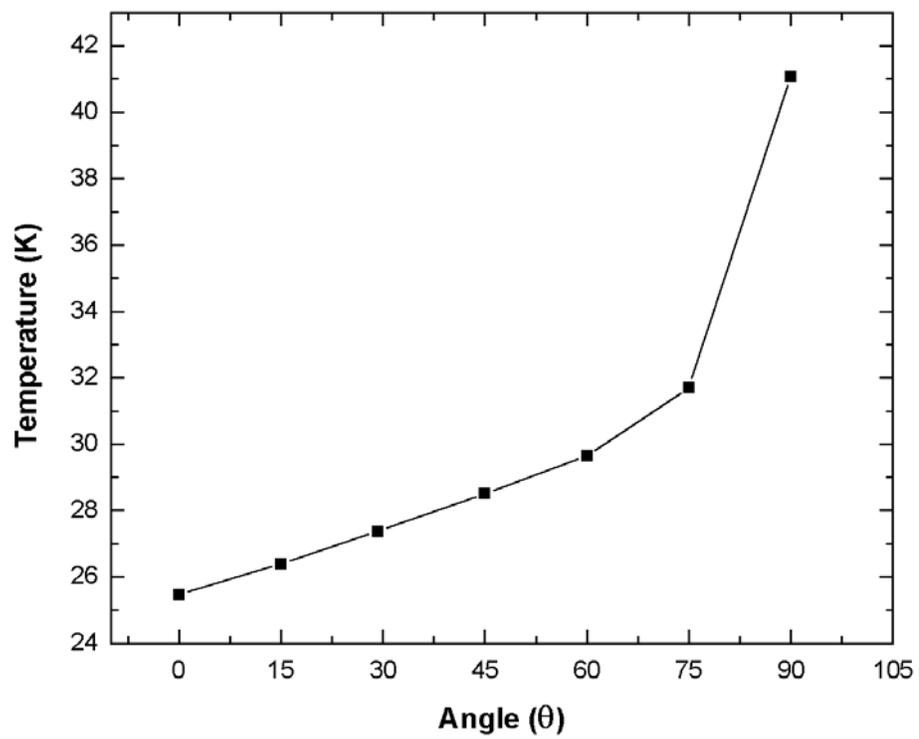

**Fig. 2**

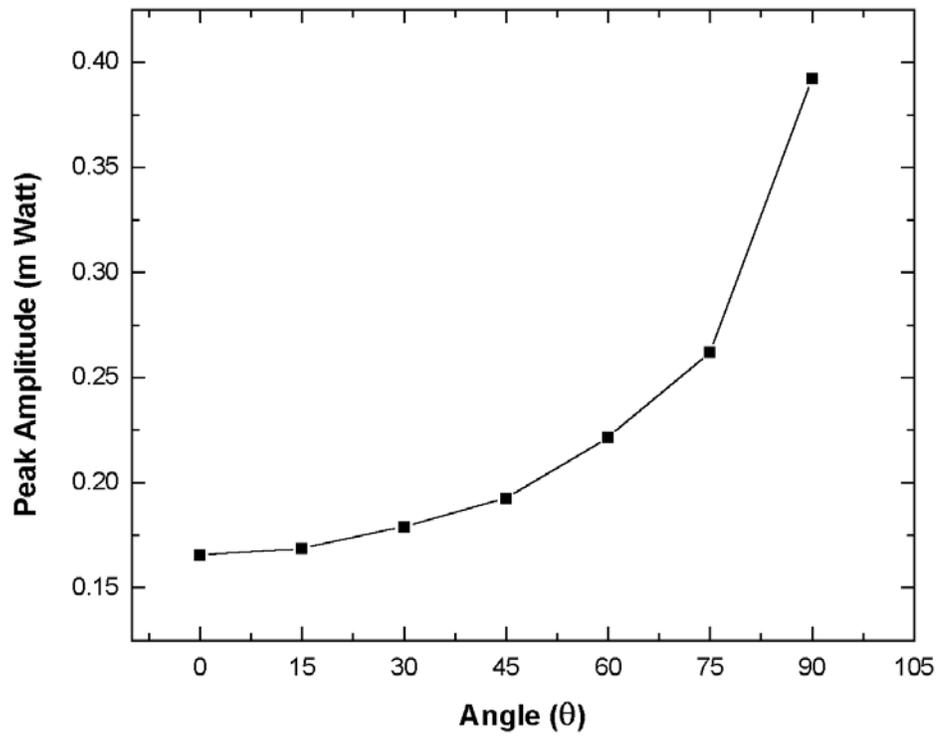

**Fig. 3**

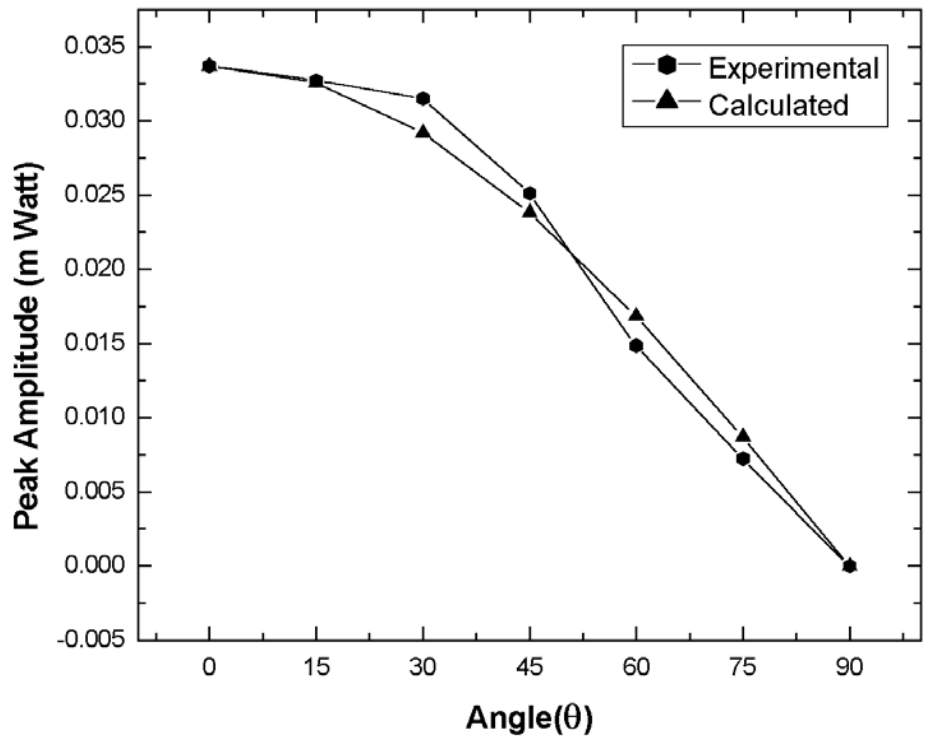

**Fig. 4**